\documentstyle[preprint,aps]{revtex}
\def\be{\begin{equation}}
\def\ee{\end{equation}}
\newcommand{\bea}{\begin{eqnarray}}
\newcommand{\eea}{\end{eqnarray}}

\def\we{\approx}
\def\fr{\frac}
\def\a{\alpha}
\def\b{\beta}

\def\d{\delta}
\def\e{\epsilon}

\def\l{\lambda}
\def\m{\mu}
\def\n{\nu}
\def\r{\rho}
\def\s{\sigma}

\def\th{\theta}
\def\w{\omega}
\def\W{\Omega}

\def\d{\delta}

\def\L{\Lambda}
\def\CL{\cal L}
\def\p{\partial}
\def\t{\tilde}
\def\nn{\noindent}
\def\no{\nonumber}
\tightenlines
\begin{document}
\title{{Hamiltonian vs Lagrangian Embedding of a Massive Spin-one
Theory Involving 2-form Field}}
\author{E. Harikumar and M. Sivakumar$^{**}$}
\address{ School of Physics,  University of Hyderabad\\
 Hyderabad, Andhra Pradesh\\
 500 046  INDIA.}
\footnotetext {$^{**}$E-mail:mssp@uohyd.ernet.in }  
\maketitle
\begin{abstract} 

We consider the Hamiltonian and Lagrangian embedding of a first-order,
massive spin-one, gauge non-invariant theory involving anti-symmetric
tensor field. We apply the BFV-BRST generalised canonical approach to
convert the model to a first class system and construct  nil-potent BFV-BRST
charge and an unitarising Hamiltonian. The canonical analysis of the
St\"uckelberg formulation of this model is presented. We bring out the
contrasting feature in the constraint structure, specifically with respect
to the reducibility aspect, of the Hamiltonian and the Lagrangian embedded
model. We show that to obtain manifestly covariant St\"uckelberg
Lagrangian from the BFV embedded Hamiltonian, phase space has to be
further enlarged and show how the reducible gauge structure emerges in the
embedded model.

\end{abstract}

\draft
\textheight 8.5in
\newpage
\begin{center}
{\bf Introduction}
\end{center}
             The problem of quantisation of gauge systems is handled
either in the covariant BRST framework \cite{brst} or in the canonical 
framework \cite{dirac,mh}.
In the BRST scheme one constructs a nil-potent BRST charge ($Q_{BRST}$)
which generates the global symmetry transformations of the gauge fixed
action and the physical states has to satisfy the condition
$Q_{BRST}|\psi>=0$. In the canonical framework, after fixing the gauge,
in the reduced phase space, Poisson brackets are replaced by Dirac brackets.
For systems with second class constraints one first replaces the Poisson
brackets with Dirac brackets, and pass over to quantum commutation
relations and treat the constraints as operator equations. But in many
cases, this procedure of Dirac runs into problems related to the factor
ordering ambiguity. To overcome the problems associated with the canonical
method,  generalised canonical quantisation methods have been developed
\cite{bfvt}. In these generalised canonical methods, the second class
system is first converted to a first class system and then schemes of
quantisation for gauge theories can be fruitfully applied. This systematic
conversion of the Hamiltonian and the second class constraints to a gauge
invariant Hamiltonian and first class constraints is called Hamiltonian
embedding. Even within the Lagrangian formulation, one can convert second
class system to a gauge invariant theory. This Lagrangian embedding is
achieved by introducing St\"uckelberg fields with compensating
transformations resulting in the gauge invariant Lagrangian starting from
a gauge non-invariant Lagrangian. In all known examples, these two
procedures are found to be equivalent.

The basic approach in Hamiltonian embedding is to enlarge the phase space
by introducing new variables using which the constraints and Hamiltonian
are modified to obtain a first class theory. In one of these schemes,
known as Batalin-Fradkin-Tyutin method, one first enlarges the phase
space using variables with same Grassman parity as that of the constraint
and modifies the second class constraints and Hamiltonian into a first
class one. Thus after embedding the model with second class algebra into a
model with involutive algebra (i.e., Poisson brackets between the
constraints vanish, at least weakly), one quantises the first class,
embedded model using well established procedures. In another variation of
generalised canonical scheme known as Batalin-Fradkin-Vilkovisky-BRST (
BFV-BRST ) quantisation, the phase space is enlarged by introducing new
phase space variables, Lagrange multipliers and their momenta
(all having the same Grassman parity as that of the second class
constraints) and also canonically conjugate fields ( this include ghost
fields) with opposite Grassman parity to that of the constraints. In this
space, a gauge invariant Hamiltonian and a nil-potent BFV-BRST charge are
obtained. Using the BFV-BRST nil-potent charge, the unitarising Hamiltonian
is constructed. The unitarity and the gauge independence are guaranteed in
this scheme of quantisation. 

Generally, the new phase space variables (with same Grassman parity as
that of the constraints) are identified with the St\"uckelberg fields in
the Lagrangian formulation. Earlier the Hamiltonian and Lagrangian
embedding analysis have been made for Proca model \cite{rb} and $2+1$
dimensional self-dual model\cite{self} and the differences in the nature
of the constraint structure of the embedded models were clarified.

In this paper we study the Hamiltonian and Lagrangian embeddings of a new
massive spin-one theory described by the Lagrangian
\be
{\CL} =- \fr{1}{4} H_{\m\n} H^{\m\n} + \fr{1}{2} G_{\m}G^{\m} + \fr{1}{2m}
\e_{\m\n\l\s}H^{\m\n}{\p^\l}{G^\s}. 
\label{nov.lag01}
\ee
This first-order formulation is different from the first-order
formulations of other well known models like Proca model, massive
Kalb-Ramond or Duffin-Kemmer-Petiau formulation by being the first-order
formulation of both Proca theory and massive Kalb-Ramond model. The
interest in the model is also due to its equivalence to topologically
massive $B{\wedge}F$ theory\cite{us}. Motivation for the study of
Hamiltonian and Lagrangian embeddings of the above model
(\ref{nov.lag01}), is due to certain novel features present here in
relation to reducibility, which are not shared by the other models studied
earlier. We employ the BFV-BRST procedure to convert this
(\ref{nov.lag01}) second class system to a first class Hamiltonian system
and also construct the nil-potent BFV-BRST charge and the unitarising
Hamiltonian. The first class constraints of the embedded system are found
to be {\it irreducible}. We then start from the Lagrangian embedded model,
which is the St\"uckelberg formulation and present its canonical analysis.
The constraint structure here is {\it reducible}. This difference in the
reducibility aspect of the constraints between the Hamiltonian and
Lagrangian embedded model is not present in the earlier models studied.
But as found in the earlier models, the difference in the nature of the
constraints modified, by these two embedding procedures is observed here
also. We also find that from the phase space path integral of the BFV
embedded model, the St\"uckelberg Lagrangian obtained is not manifestly
covariant. We show, by further enlarging the BFV-phase space, manifestly
covariant Lagrangian is obtained.  This enlargement also reconciles the
difference in the reducibility nature of the constraints in embedded
versions obtained by two procedures.

Recently the BFV-BRST procedure has been applied to few systems like the
abelian Proca model\cite{ref} and the massive super particle \cite{omer}.
The present work provides yet another illustration of this procedure,
which by itself is of intrinsic interest.

This paper is organised in the following way. In section I, we construct
the BFV-BRST nil-potent charge and unitarising Hamiltonian. In section
II, we present the Hamiltonian analysis of the Lagrangian embedded model.
In section III, we first point out the differences in the Hamiltonian
embedded model and the Lagrangian embedded model. Then using the phase
space path integral method we study the equivalence of BFV-BRST embedded
model and St\"uckelberg formulation. Here we bring out the necessity of
introducing the new phase space variables to get the manifestly covariant
St\"uckelberg Lagrangian and also clarify how the reducible nature comes
in the embedded model. We conclude with discussion in section IV.

\nn We work with $g_{\m\n}= (1,-1,-1,-1)$ and $\e_{0123}=1$.
%\newpage
\begin{center}
{\bf {I.~BFV-BRST Hamiltonian Embedding}}
\end{center}

In this section, by following Hamiltonian embedding procedure of BFV, we
systematically convert all the second class constraints of the  
first-order formulation to first class ones and construct a nil-potent
charge. Using the BFV-BRST nil-potent charge, we construct a unitarising
Hamiltonian.

We start with the Lagrangian, with the last term expressed in a 
symmetric form as,
\be
{\CL} =- \fr{1}{4} H_{\m\n} H^{\m\n} + \fr{1}{2} G_{\m}G^{\m} + \fr{1}{4m}
\e_{\m\n\l\s}H^{\m\n}{\p^\l}{G^\s} - \fr{1}{4m}\e_{\m\n\l\s}\p^\m H^{\n\l}
G^\s. 
\label{nov.lag1}
\ee
The primary constraints following from the above Lagrangian are,
\bea
\Pi_0 \we 0,~~~~~~~
\Pi_{0i}\we 0,
\label{pri.con1}
\eea
\bea
\W_i \equiv \left( \Pi_i -\fr{1}{4m}\e_{0ijk}H^{jk} \right)&\we& 0,\\ 
{\rm and}~~\L_{ij} \equiv \left ( \Pi_{ij}+\fr{1}{2m} \e_{0ijk} G^k
\right) &\we& 0. 
\label{pri.con2}
\eea
and the Hamiltonian density following from the above Lagrangian
(\ref{nov.lag1}) is, 
\bea
H_T =\fr{1}{4}H_{ij}H^{ij} -\fr{1}{2}G_i G^i + H_{0i}
\left(  \fr{1}{2}H^{0i}-\fr{1}{m} \e^{0ijk}{\p_j} G_k\right)\no\\
-\fr{1}{2} G_0 \left(G^0 +\fr{1}{m} \e^{0ijk}{\p_i}H_{jk}\right)-
(\p^i G_i)\Pi_0 +(\p^j H_{ij}) \Pi^{0i}.
\label{ham.tot}
\eea
The persistence of the primary constraints leads to the
following secondary constraints,
\bea
\L \equiv \left( G_0 + \fr{1}{2m}\e_{0ijk}{\p^i}H^{jk} \right) \we 0,\no\\
\L_i \equiv \left(-H_{oi} + \fr{1}{m}\e_{0ijk}{\p^j}G^k \right) \we 0.
\label{sec.con}
\eea
%
%The non-vanishing Poisson brackets between these linearly independent 
%constraints are
%%
%\bea
%\left\{ \Pi_{0}({\vec x}) , \L({\vec y})\right\} = -\delta({\vec x -\vec
%y}),\no\\ 
%\left\{ \Pi_{0i}({\vec x}) , \L^{j}({\vec y})\right\}= \delta^{j}_{i}
%\delta({\vec x -\vec y}),\no\\ 
%\left\{ \W_{i}({\vec x}) , \L_{j}({\vec y})\right\} =
%\fr{1}{m}\e_{0ijk}\p^k\delta({\vec x -\vec y}),\no\\ 
%\left\{ \W_{i}({\vec x}) ,\L_{jk}({\vec y})\right\} =
%-\fr{1}{m}\e_{oijk}\delta({\vec x -\vec y}),\no\\ 
%\left\{ \L_{ij}({\vec x}), \L({\vec y})\right\} =
%\fr{1}{m}\e_{oijk}\p^k\delta({\vec x -\vec y}). 
%\label{non.pb}
%\eea
%%
It can be easily seen from their Poisson brackets that all the constraints
are second class as expected for a theory without 
any gauge invariance. Note that the constraints $\W_i~~{\rm and}~~ \L_{ij}$
are due to the symplectic structure of the Lagrangian (\ref{nov.lag1}).
Following Faddeev and Jackiw \cite{fj}, the symplectic conditions, which
are not true constraints, are implemented strongly leading to the 
modified bracket, 
\be
\left\{ G_{i}({\vec x}),H_{jk}({\vec y})\right\}=-m\e_{0ijk}
\delta({\vec x -\vec y}).
\label{sympl}
\ee
Consequently, $\W_i$ and $\L_{ij}$ are implemented strongly.

Following the generalised Hamiltonian method, we enlarge the phase space
by introducing canonically conjugate pair of ghost fields $C(x)_\a$ and 
$P(x)^\a$ corresponding to each constraint satisfying,
\bea
\{C(x)_\a,P(y)^\b\}&=&i\hbar \d_{\a}^\b \d(x-y),\no\\
gh(C_\a)=-gh(P^\a)&=&1,~~~
{\cal E} (C_\a)={\cal E}(P_\a)=1,~~~~\a=1,2,3,4.
\label{bfv1}
\eea
\nn where $gh~{\rm and}~{\cal E}$ stand for the ghost number and
the  Grassman parity respectively and $\{,\}$ stands for the graded
commutator\footnote{Following the BFV-BRST construction we replace all the
Poisson brackets between the variables of the original second class model
with commutators.}.

Now we define the operators,
\bea
\W ={\t T}_\a C^\a +\Sigma_{n=1}^{\infty} \fr{1}{(n+1)!n!}
P_{\a{1}}....P_{\a{n}}
U_{\b{n+1}~\b{n}...\b{1}}^{\a{1}~\a{2}....\a{n}}C^{\b{1}~\b{2}.....\b{n+1}}
\no\\
\W^\a= V_{\b}^\a C^\b +\Sigma_{n=1}^{\infty} \fr{1}{(n+1)!n!} 
P_{\a{1}}....P_{\a{n}}
V_{\b{n+1}~\b{n}...\b{1}}^{\a \a{1}~\a{2}....\a{n}}C^{\b{1}~\b{2}.....\b{n+1}},
\label{bfv2}
\eea
satisfying the conditions,
\bea
\{\W(x), \W(y)\}=\int dx dy \W^\a \w_{\a\b}\W^\b\no\\
\{\W^\a, \W^\b\}=0,~~~~
\{\W^\a, \W\}=0.
\label{bfv.con}
\eea
where ${\t T}_\a~{\rm for}~ \a=1,2,3,4$ are the original second class 
constraints $\Pi_0,~\Pi_{0i},~ \L,~{\rm and}~\L_i$
({\ref{pri.con1}},{\ref{sec.con}}) respectively and $\w_{\a\b}$ is an
arbitrary, invertable anti-symmetric matrix. In our case we find 
\bea
\W &=&\Pi_o C^1 +\Pi_{0i}C^{2i} +\L C^3 +\L_iC^{4i}\label{bfvop}\\
\W_1&=&\fr{1}{\sqrt 2} (C_1 -C_3),~~~~~~ 
\W_2= \fr{1}{\sqrt 2}(C_{2}^i +C_{4}^i),
\no\\
\W_3&=&\fr{1}{\sqrt 2}(C_1+C_3),~~~~~~ \W_4=\fr{1}{\sqrt 2} (C_{2}^i -C_{4}^i),
\label{bfvop1}
\eea
satisfying the above conditions (\ref{bfv.con}). 

We now further enlarge the phase space by introducing the
variables $\Pi_\a,~\a, p_i,~{\rm and}~ q_i$ obeying
\bea
\{\a(x), \Pi(y)_\a\}= i\hbar \d(x-y)\no\\
\{q(x)_i, p(y)^j\}= i\hbar \d_{i}^j\d(x-y),\no\\
gh(\a)=gh(\Pi_\a)=gh(p_i)=gh(q_i)=0,\no\\
{\cal E}(\a)={\cal E}(\Pi_\a)={\cal E}(p_i)={\cal E}(q_i)=0,
\label{bfv.pb}
\eea
and define a nil-potent operator 
\bea
{\W}^{\prime} &=&\W -\W_1 \Pi_\a -\W_{2}^i p_i + \W_3 \a +\W_{4}^i q_i
\no\\
&=&T_\a C^\a,\no\\
{\rm and} ~~~~~~(\W^{\prime})^2& =&\fr{1}{2}\{\W^{\prime},\W^{\prime}\}=0.
\label{bfv.nil}
\eea
Here
\bea
T_1 &=& \Pi_0 -\fr{1}{\sqrt 2} (\pi_\a -\a),~~~~~
T_{2}^i = \Pi^{0i} -\fr{1}{\sqrt 2} (p^i +q^i),\no\\
T_3 &=& \L +\fr{1}{\sqrt 2} (\pi_\a +\a),~~~~~
T_{4}^i =\L^i -\fr{1}{\sqrt 2} (p^i -q^i),
\label{bfv.in}
\eea
are the modified constraints that are in strong involution. Note that
these first class constraints are linearly independent which can be seen
by the degrees of freedom count. 

To construct a first class Hamiltonian that is in strong involution
with $T_\a$, we first define,
\bea
{\bar \W}_1 &=&\fr{1}{\sqrt 2} (P_1 -P_3),~~~~~~~~~~~~~
{\bar \W}_{2}^i= \fr{1}{\sqrt 2} (P_{2}^i +P_{4}^i),\no\\
{\bar \W}_3&=&\fr{1}{\sqrt 2}(P_1+P_3),~~~~~~~~~~~~~~~
{\bar \W}_4=\fr{1}{\sqrt 2} (P_{2}^i -P_{4}^i),
\label{bfv.op1}
\eea
satisfying the conditions,
\bea
\{\W(x)_\a, {\bar \W(y)}^\b\} &=&i\hbar \d_{\a}^\b \d(x-y),\no\\
{\bar \W}_\a \W^\a &=&P_\a C^\a.
\label{bfv.op2}
\eea

Using these operators, one defines the first class Hamiltonian,
\bea
{\cal H} &=& H_T +\fr{1}{(i\hbar)^2}\int dx \{H_T, \{\W,{\bar \W}^\a\}\}
\Phi_\a \no\\
&+&\fr{1}{(i\hbar)^4} \int dx dy \{\{H_T, \{ \W, {\bar \W}^\a\}\}, 
\{\W, {\bar \W}^\b\}\}\Phi_\a \Phi_\b +....,
\label{bfv.ham1}
\eea
where $\Phi_\a$ is $ \a, \Pi_\a, p_i, ~{\rm and}~ q_i$ for $\a=1,2,3 
~{\rm and}~ 4$ respectively and the total Hamiltonian $H_T$ is given in
(\ref{ham.tot}). We see that all the higher order terms in the above series
vanish. 

Using (\ref{ham.tot},\ref{bfvop},\ref{bfv.op1},\ref{bfv.op2}) in
(\ref{bfv.ham1}), we get 
\bea
{\cal H} &=& H_T -\L \Pi_\th +\p^i \Pi_0 \p_i \th -\L_i {\t \Pi}^i +
(\p^j \p_j \Pi_{0i} -\p^j\p_i \Pi_{0j})B^i\no\\
&-& \fr{1}{2} (\Pi_\th)^2 +\fr{1}{2} {\p^i \th} {\p_i \th}
+\fr{1}{2} {\t \Pi}_i {\t \Pi}^i -\fr{1}{4} B_{ij}B^{ij},
\label{bfv.ham}
\eea
where we have used the definitions,
\bea
\fr{1}{\sqrt 2} (\a -\Pi_\a) &=& \th,~~~
\fr{1}{\sqrt 2} (\a +\Pi_\a) = \Pi_\th,\no\\
\fr{1}{\sqrt 2}( q_i -p_i) &= &B_i,~~~~
\fr{1}{\sqrt 2} (q_i + p_i) = {\t \Pi}_i,
\label{red.bfv}
\eea
and $B_{ij} =(\p_i B_j -\p_j B_i)$. From (\ref{red.bfv}) we notice that
$\{\th(x), \Pi(y)_\th\} =i\hbar \d(x-y)~{\rm and}~ \{B(x)_i, {\t
\Pi(y)}^j\}=i\hbar \d_{i}^j \d(x-y).$ Thus, we can re-express the
involutive constraints (\ref{bfv.in}) as
\bea
T_1 &=& \Pi_0 +\th,~~~~~~~~T_{2}^i=\Pi^{0i} +B^i,\no\\
T_3 &=& \L +\Pi_\th,~~~~~~~T_{4}^i =\L^i- {\t \Pi}^i.
\label{red.in}
\eea
Now we have the modified constraints $T_\a$ , Hamiltonian ${\cal H}$
and the nil-potent operator $\W^{\prime}$ satisfying
\bea
\{T_\a, T_\b\}&=&0,\no\\
\{T_\a, {\cal H}\}&=&0,\no\\
{\rm and}~~~\{\W^{\prime}, {\cal H}\}&=&0.
\label{bfv1.al}
\eea

Now we further enlarge the phase space by introducing (dynamical) Lagrange
multipliers $\l_\a$ and their conjugate momenta $\Pi_{\l}^\a$  corresponding 
to each of these first class constraints $T_\a$ satisfying,
\bea
\{\l(x)_\a, \Pi(y)_{\l}^\b\}&=&i\hbar \d_{\a}^\b \d(x-y),\no\\
gh(\l_\a) = gh (\Pi_{\l a}) &=& 0,~~~~~
{\cal E}(\l_\a) ={\cal E}(\Pi_{\l a})= 0.
\label{bfv.mul}
\eea
We also introduce the hermitian ghost-anti ghost pairs,
\be
\{g(x)_\a, {\bar g(y)}^\b\} =i\hbar \d_{\a}^\b \d(x-y),
~~~gh(g_\a)=-gh({\bar g}_\a) =1,
~~~~{\cal E}(g_\a) = {\cal E}({\bar g}_\a) = 1
\label{bfv.gh}
\ee
and define the BFV-BRST operator
\be
Q= \W^{\prime} +\int dx \Pi_{\l}^\a g_\a
\label{bfv.brst1}
\ee
which is nil-potent  $(Q^2 =0)$.

\nn The gauge fixing Fermion is defined as,
\be
\Psi = \int dx [ P_\a \l^\a + {\bar g}^\a \chi_\a],~~~~~~~~~gh(\Psi)=-1,
\label{bfv.gff}
\ee
where  $\chi_\a$ are the gauge fixing conditions to be fixed so that 
$det \{T_\a,\chi_\b\}\ne0.$ Using this $\Psi,$ we construct the unitarising 
Hamiltonian
\be
{\cal H}_u = {\cal H} + \fr{1}{i\hbar} \{\Psi, Q\}.
\label{uni.ham}
\ee
This completes the application of the BFV-BRST procedure to (\ref{nov.lag1}).

\begin{center}
{\bf{II.~Lagrangian Embedding}}
\end{center}
In  this section we present the Hamiltonian and the constraints 
the embedded Lagrangian. We start with the St\"uckelberg Lagrangian
corresponding to (\ref{nov.lag1}) is,
\bea
{\cal L}&=&-\fr{1}{4}(H_{\m\n}-B_{\m\n})(H^{\m\n}-B^{\m\n})
+\fr{1}{2}(G_\m+\p_\m \th)(G^\m +\p^\m \th)\no\\
&+& \fr{1}{4m}
\e_{\m\n\l\s}H^{\m\n}{\p^\l}{G^\s} - \fr{1}{4m}\e_{\m\n\l\s}\p^\m H^{\n\l} 
G^\s,
\label{st.nlag}
\eea
where $B_{\m\n}=(\p_\m B_\n -\p_\n B_\m)$. The above Lagrangian is
invariant under
\bea
\d H_{\m\n}&=&(\p_\m \L_\n -\p_\n \L_\m),
~~\d B_\m = \L_\m +\p_\m \a,\no\\
\d G_\m &=&\p_\m \l,~~~~~~~
\d \th = \l.
\label{sym.st}
\eea
The Lagrangian is invariant under the variation of the gauge 
parameter $\L_\m~(\d \L_\m = \p_\m\r)$, making the system reducible.

Next the canonical analysis of the gauge invariant theory
described by the above St\"uckelberg  Lagrangian is carried out to compare
with the Hamiltonian and constraint structure of the embedded model.
%bringing out their contrasting features.

The primary constraints following from the Lagrangian (\ref{st.nlag}) are,
\bea
\Pi_0&\we&0,~~~\Pi_{0i}\we0,\\
{\t \Pi}_0&\we&0,
\label{prist.con}
\eea
\be
\W_i=(\Pi_i +\fr{1}{4m}\e_{0ijk}H^{jk})\we0,~~
\L_{ij}=(\Pi_{ij}+\fr{1}{2m}\e_{0ijk}G^k)\we0,
\label{prist.con1}
\ee
and the Hamiltonian is,
\bea
H_{St}&=& \fr{1}{2} (\Pi_\th)^2 -\fr{1}{2}{\t \Pi}_i{\t \Pi}^i +\fr{1}{4}
(H_{ij}-B_{ij})(H^{ij}-B^{ij}) -\fr{1}{2}(G_i+\p_i\th)(G^i +\p^i\th)\no\\
&+& H^{0i}({\t \Pi}_i -\fr{1}{m}\e_{0ijk}\p^jG^k) 
-G_0(\Pi_\th +\fr{1}{2m}\e_{0ijk}\p^i H^{jk})-B_0(\p^i{\t \Pi}_i),
\label{st.ham}
\eea
and the Gauss law constraints are,
\bea
\L&=&(\Pi_\th +\fr{1}{2m}\e_{0ijk}\p^iH^{jk})\we0,\\
\L_i&=&-({\t \Pi}_i -\fr{1}{m}\e_{0ijk}\p^jG^k)\we0,\\
\w&=&\p^i{\t \Pi}_i\we0.
\label{st.gauss}
\eea

Here $~\Pi_\m,~\Pi_{\m\n},~{\t\Pi}_\m ~{\rm and}~\Pi_\th~$ are the conjugate
momenta corresponding to $G^\m, H^{\m\n},$ $ B^\m,~{\rm and}~\th$.
Imposing the symplectic conditions (\ref{prist.con1}), results in the same
modified bracket as in (\ref{sympl}). All the above constraints are first
class as one can  verify from their Poisson bracket algebra. The
reducibility of the model is evident from the fact that the constraints
$\L_i~{\rm and}~\w$ are linearly dependent $(\p^i\L_i + \w =0)$. Thus we
have nine linearly independent   first class constraints and hence the
model defined by (\ref{st.nlag}), describes massive spin-one particles. 

\begin{center}
{\bf {III.~ Hamiltonian Embedding vs Lagrangian Embedding}}
\end{center}

In this section we make a comparitive analysis of the constraint structure
of BFV embedded model and the one following from the St\"uckelberg
formulation (\ref{st.nlag}). Note the following differences in the
constraint structure of the Hamiltonian and Lagrangian embedded
versions of this model (\ref{nov.lag1}). 
\begin{enumerate}
\item{The Hamiltonian embedding
procedure modifies all the second class constraints of the original model
whereas it is observed that the St\"uckelberg formulation modifies only
the Gauss law constraints and not the primary constraints (This feature is
present in the case Proca model also\cite{rb}).}
\item{The Hamiltonian embedded model has only {\it irreducible} first
class constraints while the constraints of the Lagrangian embedded model
obeys a {\it reducible} gauge algebra.}
\end{enumerate}
The first apparent difference between these two embeddings can be 
accounted by rewriting the Lagrangian (\ref{st.nlag}) by dropping surface
terms (i.e, by re expressing $\fr{1}{2}(H_{\m\n}B^{\m\n})$ as $-(\p^\m
H_{\m\n}B^\n)$ and $G_\m \p^\m \th$ as $-(\p^\m G_\m \th)$). With the
Lagrangian written in this form, we get both the primary and the secondary
constraints of the St\"uckelberg formulation which are structurally
modified. Then the constraints will be in the same footing as the BFV
embedded system. 

Next, the second aspect, the difference in the reducible nature is
clarified by studying the equivalence of the BFV embedded model to the
St\"uckelberg formulation using the phase space path integral approach. 
For this, we start with the BFV embedded model partition function,
\be
Z =\int D\eta~ exp~i\int d^4 x L, 
\ee
where the meassure is,
$$
D\eta=D\Pi_0 DG_0 DG_i D\Pi_{0i}DH_{0i} DH_{ij} D\Pi_\th D\th 
D{\t \Pi}_i DB_iD\l_\a D\Pi_{\l}^\a DP_\a DC_\a Dg_\a D{\bar g}_\a
$$ and the $L={\cal P}\p^0 {\cal Q} -{\cal H}_u,$ ${\cal P}$ is the generic
momentum, and ${\cal Q}$ is the generic field. Here ${\cal H}_U$ is the
unitarising Hamiltonian given in ({\ref{uni.ham}).

To show this equivalence, we chose the gauge fixing conditions to be
\be
\d (\chi_{1}) = \d(G_0),~~~~~\d(\chi_{2})= \d(H_{0i}),
\label{st.gfc1}
\ee
\be
\d(\chi_{3})= \d (\p_iG^i),
~~~~~\d(\chi_{4}^{ij})=\d( H^{ij}-\fr{1}{m}\e^{0ijk}\p_k G_0).
\label{st.gfc2}
\ee
With this choise of $\chi_\a$, we get
\be
\fr{1}{i\hbar}\{\Psi,Q\}= -\l_\a T^\a +P_\a g^\a -\chi_\a \Pi_{\l}^\a
+{\bar g}_{\a} C^{\b} \{\chi^{\a}, T_{\b}\}.
\ee

The integrations over ${\bar g}_\a,$ and $C_\a$ give the Faddeev-Popov
determinant $det|\{\chi^{\a}, T_{\b}\}|$, which is trivial in our case and
we drop it hereafter. The integrations over $ g_\a,$  and $P_\a$ give
only constant numerical factors (which we omit in the following) and that
of $ \Pi_{\l}^\a$ and $\l_\a$ gives $\d(T_\a)$ and $\d(\chi_\a)$ respectively.

\nn Thus, the partition function becomes
\be
Z =\int D\Pi_0 DG_0 DG_i D\Pi_{0i}DH_{0i} DH_{ij} D\Pi_\th D\th 
D{\t \Pi}_i DB_i \d(T_\a) \d(\chi_\a)~exp~i\int d^4 x {\cal L},
\label{bfv.pf1}
\ee
where, $T_\a$ are the first class constraints (\ref{red.in}), and
\bea
{\cal L}&=& \Pi_0 \p^0G^0 +\Pi_{0i}\p^0 H^{0i} +{\t \Pi}_i\p^0 B^i 
+\Pi_0\p^0\th \no\\
&+&\fr{1}{4m}\e_{0ijk}H^{ij}\p^0 G^k
-\fr{1}{4m}\e_{0ijk}G^i \p^0 H^{jk} -H.
\label{bfv.pf2}
\eea
Here, 
\be
H = {\cal H} +(\L +\Pi_\th) \Pi_\th +(\L_i -{\t \Pi}_i){\t \Pi}^i.
\label{hpf}
\ee
where ${\cal H}$ is the gauge invariant Hamiltonian (\ref{bfv.ham}) and
$H$ differs from it only by terms proportional to the first class constraints
and hence both these Hamiltonians define the same gauge system (i.e., on the
constraint surface $H={\cal H}).$ 

After exponentiating the constraints $\d(\L+\Pi_\th)~{\rm and}~\d(\L_i
-{\t \pi}_i)$ using the Fourier transform fields $ c$ and $d_i$ respectively,
we integrate out $\Pi_0~{\rm and}~\Pi_{0i}$ which are
trivial due to the presence of the constraints $\d(\Pi_0 +\th)~{\rm and}~
\d(\Pi_{0i}-{\t \Pi}_i)$. Similarly the $G_0~{\rm and}~H_{0i}$ integrations
are trivial because of the gauge fixing conditions (\ref{st.gfc2}).
Redefining $c$ and $d_i$ as $G_0$ and $H_{0i}$ after integrating out
$\Pi_\th$ and ${\t\Pi}_i$, and using $\d(\chi_3)$ and the condition $\p^i
H_{ij}=0$ implied by $\d(\chi_{4}^{ij}),$ the partition function becomes
\be
Z =\int DG_\m DH_{\m\n} D\th DB_i \d(\chi_3)\d(\chi_{4}^{ij})~exp~i\int
d^4 x {\cal L}, 
\label{bfv.pf11}
\ee
where
\bea
{\cal L}&=& \fr{1}{4m}\e_{\m\n\l\s}H^{\m\n}G^{\l\s}
+\fr{1}{2}(G_\m+\p_\m\th)(G^\m+\p^\m\th) -\fr{1}{4}(H_{ij}-B_{ij})
(H^{ij}-B^{ij})\no\\
&-&\fr{1}{2}H_{0i}H^{0i} + H_{0i}\p^0B^i -\fr{1}{2}(\p^0B_i)(\p_0B^i).
\label{bfv.st1}
\eea

It should be noted that the Lagrangian (\ref{bfv.st1}) is not manifestly
covariant. This is in contrast with the other models like self-dual model
in $2+1$ dimensions, Proca model in $3+1$ dimensions where the embedding
procedure has been applied, where using the phase space path integral
approach the corresponding St\"uckelberg Lagrangians have been
obtained\cite{rb} in a manifestly covariant way. It should be noted here
that $B_0$, the time component of the vector St\"uckelberg field $B_\m$
does not appear in the embedded model. To obtain the manifestly covariant
St\"uckelberg Lagrangian (\ref{st.nlag}), we further enlarge the phase
space of the embedded model by introducing a pair of conjugate variables 
${\t \Pi}_0~{\rm and}~ B_0$ along with certain constraints such that the
degrees of freedom is not changed. A natural choice is to introduce 
${\t\Pi_0=0}$
as a first class constraint and $B_0=0$ as the corresponding gauge fixing
condition and these conditions will remove the two extra degrees of
freedom we have introduced $(B_0~{\rm and}~ {\t \Pi}_0)$.
Thus the Gauss law constraint generated by demanding the consistency
of ${\t \Pi}_0\we0$ has to be
linearly dependent with the other first class constraints.
We include the linearly independent constraint $\d ({\t \Pi}_0)$ and the
gauge fixing condition $\d(B_0)$ in the path integral
measure and $B_0\p_i({\t \Pi}_i + H_{0i})$ in the exponential along with 
the integrations over $B_0$ and ${\t \Pi}_i$ in eqn (\ref{bfv.pf1}). Here
the term $B_0\p_i({\t \Pi}_i + H_{0i})$ introduced in the path
integral, will give the Gauss law constraint when we demand the
persistency of  ${\t \Pi}_0$, and this turn out to be a first class
constraint which is linearly dependent with already existing first class
constraint $T_{4}^i~ (\ref{red.in})(viz: \p_i T_{4}^i + \p_i({\t \Pi}_i +
H_{0i})=0)$. Next using the constraint $\d({\t\Pi}_0)$ we carry out the
integration over ${\t \Pi}_0$ dropping all ${\t \Pi}_0$ dependent terms.
With the gauge condition  $\d(B_0)$ and integration over $B_0$ in the
partition  function, we get the Lagrangian in the exponential which is
same as (\ref{st.nlag}), showing the gauge equivalence of  BFV embedded
model to the St\"uckelberg formulation. Thus we see that the enlargement of
the phase space of BFV embedded model is essential in obtaining the manifestly
covariant St\"uckelberg action and it also brings out  the reducible
nature of the constraints. 

In the earlier examples where the embedding had been applied
\cite{rb,ref,ref1,ref2,ref3}, the number of the new phase space
variables used to modify the original  second class constraints to first
class and the number of the phase space variables corresponding to the
St\"uckelberg fields were the same. But in the case of the model studied
in this paper, the  
embedded model has two phase space variables less than that of the  
St\"uckelberg formulation. It is interesting to note that this
extra two variables and the associated constraints required to obtain the 
manifestly covariant St\"uckelberg Lagrangian starting from the phase 
space partition function of the embedded model also make the theory reducible.
This seems to indicate a deep link between the manifest covariance of the 
Lagrangian and the reducible gauge structure.

\begin{center}
{\bf{IV. Conclusion}}
\end{center}

In this paper, we have made a comparative analysis of Hamiltonian and
Lagrangian embedding of a first-order formulation of massive spin-one
theory. Following the generalised Hamiltonian procedure of BFV, we have
converted a second class system to a first class one and the nil-potent 
BFV-BRST charge and unitarising Hamiltonian were constructed. The
canonical analysis of the gauge invariant St\"uckelberg 
Lagrangian corresponding to the initial gauge non-invariant theory was
presented. We have then studied the relation between the Hamiltonian and
Lagrangian embedding procedures. We find the first class constraints
obtained by Hamiltonian embedding are irreducible but the ones obtained
from Lagrangian embedding are reducible. Also the Lagrangian obtained from
the BFV-Hamiltonian using the phase space path integral by integrating out
all the momenta variables is not manifestly covariant. By further
enlarging the BFV phase space  by introducing new canonically conjugate
variables with new constraints, manifestly covariant St\"uckelberg
Lagrangian can be obtained starting from  the phase space partition
function of the embedded model. It turns out that one of the new
constraints (Gauss law constraint) is linearly dependent on already
existing constraints leading to reducibility. Thus this explains the
difference in the reducible nature of the constraints in the Hamiltonian
formulation of the St\"uckelberg theory and embedded model. It also seems
to imply a relationship between obtaining  manifest covariance and
reducibility of the constraints. A similar feature was also observed in a
dually symmetric Maxwell theory where reducibility ( of second class
constraints ) occur in the manifestly covariant formulation\cite{rab}.

\vskip0.5cm

\noindent {\bf Acknowledgements}:
We thank R. Banerjee for useful comments. EH thanks U.G.C., India for
support through S.R.F scheme.

\end{document}